\documentclass[prb,twocolumn,superscriptaddress,showpacs,amsmath,amssymb]{revtex4}
\usepackage{amsmath}
\usepackage{amssymb}
\usepackage{bm}
\usepackage{graphicx}

\DeclareMathAlphabet{\bi}{OML}{cmm}{b}{it}

\begin{document}
\title{Plasmons and their interaction with electrons in trilayer graphene}
\author{P. M. Krstaji\'c}
\affiliation{Institute of Microelectronic Technologies and Single Crystals (IHTM), University of Belgrade,
Njego\v{s}eva 12, 11000 Belgrade, Serbia}
\affiliation{Departement Fysica, Universiteit Antwerpen, Groenenborgerlaan 171, B-2020 Antwerpen, Belgium}
\author{B. Van Duppen}
\affiliation{Departement Fysica, Universiteit Antwerpen, Groenenborgerlaan 171, B-2020 Antwerpen, Belgium}
\author{F. M. Peeters}
\affiliation{Departement Fysica, Universiteit Antwerpen, Groenenborgerlaan 171, B-2020 Antwerpen, Belgium}

\begin{abstract}
The interaction between electrons and plasmons in trilayer graphene is investigated within the Overhauser approach resulting in the 'plasmaron'
quasi-particle. This interaction is cast into a field theoretical
problem, and its effect on the energy spectrum is calculated using improved Wigner-Brillouin perturbation theory.
The plasmaron spectrum is shifted with respect to the bare electron spectrum by $\Delta E(\mathbf{k})\sim 150\div200\,{\rm meV}$
for \textit{ABC} stacked trilayer graphene and for \textit{ABA} trilayer graphene by $\Delta E(\mathbf{k})\sim 30\div150\,{\rm meV}$
($\Delta E(\mathbf{k})\sim 1\div5\,{\rm meV}$) for the hyperbolic (linear) part of the spectrum.
The shift in general increases with the electron concentration $n_{e}$ and electron momentum.
The dispersion of plasmarons is more pronounced in \textit{ABC} stacked than in \textit{ABA} stacked trilayer graphene,
 because of the different energy band structure and their different plasmon dispersion.
\end{abstract}
\pacs{73.22.Pr, 73.20.Mf, 71.10.-w}
\maketitle

\vspace{5mm}
\section{Introduction}\label{intro}
Trilayer graphene as a novel material has attracted considerable attention in recent years\cite{trilayerNat,Zhang1,Ben1,ABC1,ABC2}. Trilayer and few layer graphene are
interesting because they possess different and unique properties with respect to both
single layer graphene and conventional semiconductors. For instance, bilayer\cite{bilgap} and certain types of trilayer graphene\cite{trilayerNat}
are shown to have an electrically tunable band gap\cite{Peet1}. This would make them good candidates for application in electronic industry where the control of
the band gap is desirable to implement a chosen logic. On the other hand, plasma excitations in graphene structures\cite{graphplasm}
and in general nanostructures made of graphene, semiconductors and/or metals are the subject of current interest of many researchers worldwide.
The new emerging field called plasmonics\cite{Maier} and nanoplasmonics is concerned with the methods to confine plasmons and electromagnetic fields
over dimensions on the order of or smaller than its wavelength. For instance, surface plasmons guided by graphene structures are shown to exhibit
low losses and being tunable by gating and doping makes graphene an appropriate candidate to replace metal plasmonic devices\cite{graphplasm,Junanot}.

In order to investigate the electron-plasmon interaction in more detail, the concept of
a new quasiparticle named "plasmaron" was introduced which is in fact a bound state of a charge carrier with plasmons. Coulomb interaction and plasmarons
in both single layer graphene\cite{plasmaron1,plasmaron2,plasmaron3,plasmaron_science,fpPark}
and bilayer graphene\cite{Sabas,Sensarma} have been studied intensively. One of the reasons is that it was found experimentally that,
in monolayer graphene\cite{plasmaron_science} the accepted view of a linear
(Dirac-like) spectrum does not provide a sufficiently detailed picture of the charge carrying excitations in this material.
 The motivation behind the interest in this kind of studies is that studying the physics of the interaction
between electrons and plasmons may lead to the realization of plasmonic devices
which merge photonics and electronics. The interest in similar phenomena in few layer graphene is equally high.

Coulomb interaction and electronic screening was probed in bilayer and multilayer graphene using angle-resolved photoemission
spectroscopy (ARPES) in Ref.~\onlinecite{BostwickPRL}. Further plasmon dispersion was studied in multilayer graphene using high-resolution
electron energy-loss spectroscopy in Ref.~\onlinecite{plasmondisp}.
Recently, plasmarons and the quantum spectral function in bilayer graphene
was investigated in Ref.~\onlinecite{Sensarma} theoretically, where the onset of a broad plasmaron peak away from
the Fermi surface was predicted.  While
the energy dispersion in graphene is linear in momentum, in trilayer graphene it can be cubic, hyperbolic and/or
linear depending on the stacking order. The advantage of multilayer
graphene over usual semiconductors is that its charge carrier density can be controlled by the application of a gate voltage
over orders of magnitude and the charge carrier type can be changed from electrons to holes.
Furthermore, the band gap can be tuned to meet specific demands for device design.

In this work, we use second order perturbation theory in order to take into account the electron-plasmon
interaction which is cast into a field theoretical problem. In this way, one is able to calculate the correction
to the band structure which comes as a consequence of the interaction of charge carriers with plasmons. The interaction is treated using
the Overhauser approach\cite{Overhauser,remotepol} here applied to the two-dimensional electron gas in trilayer graphene.

The paper is organized as follows. In Sec.~\ref{theory} we present the theoretical model and derive the relevant expressions for the interaction and
the coupling between electrons and plasmons in trilayer graphene. In the subsequent section, Sec.~\ref{results}, the numerical calculations
of the energy correction due to the interaction with plasmons are presented as a function of electron momentum and for
various doping levels, i.e. charge carrier density. Both stacking order \textit{ABC} and \textit{ABA} were considered. The influence of the doping level is analyzed and discussed.
Finally, we summarize our results and present the conclusions in Sec.~\ref{summary}.

\section{Theoretical model}\label{theory}
\subsection{\textit{ABC} stacked trilayer graphene}
If the relevant energies of interest in trilayer graphene are smaller than the interlayer hopping parameter $\gamma_{1}$, one may use the low energy limit.
In this limit, the problem can be reduced to the effective two-band model and the corresponding Hamiltonian reads\cite{Kotov}
\begin{equation}\label{Ham0}
H_{ABC}=\frac{(\hbar v_{F})^{3}}{\gamma_{1}^{2}}\left[\begin{array}{cc}
0 & (k_{x}-ik_{y})^{3} \\
(k_{x}+ik_{y})^{3} & 0 \end{array}\right]\,,
\end{equation}
\noindent where $v_{F}$ is the Fermi velocity. The eigenvalues of Eq.~(\ref{Ham0}) are known and read
\begin{equation}\label{eigenener}
E^{(0)}_{l}=l\frac{(\hbar v_{F})^{3}}{\gamma_{1}^{2}}k^{3}\,,
\end{equation}
\noindent where $l=\pm1$.
Structures made of graphene may exhibit and support quanta of collective charge excitations of the electron gas,
i.e. plasmons, as a result of the restoring force of the long-range Coulomb interaction.
Contrary to the case of conventional two-dimensional electron gas, the "Dirac plasma" is manifestly of quantum nature\cite{SarmaPRL}.
For example, in single layer graphene the plasma frequency is proportional to $1/\sqrt{\hbar}$, and does not have a classical limit independent of the Planck constant.
The dynamic dielectric function within Random Phase Approximation (RPA) is given by

\begin{equation}\label{dielPi}
\epsilon(\mathbf{q},\omega)=1-\frac{2\pi e^{2}}{\kappa q}\Pi(\mathbf{q},\omega)\,,
\end{equation}
\noindent where $\kappa=(1+\kappa_{s})/2$ is the dielectric constant of the material and is related to the one of the substrate.
Here $\Pi(\mathbf{q},\omega)$ is the free-particle polarizability and is given by
 \begin{equation}
\Pi(\mathbf{q},\omega)=\frac{g_{d}}{\Omega}\sum_{ll^{\prime}\mathbf{k}}
\frac{f_{\mathbf{k}}^{l}-f_{\mathbf{k+q}}^{l^{\prime}}}{\hbar\omega+E_{\mathbf{k}}^{l}-E_{\mathbf{k+q}}^{l^{\prime}}}
F_{ll^{\prime}}(\mathbf{k},\mathbf{k+q})\,,
\end{equation}
\noindent where $g_{d}$ is the degeneracy factor, $f_{\mathbf{k}}^{l}$ is the Fermi-Dirac distribution, and $\Omega$ is the volume of the system.
Here $F_{ll^{\prime}}$ is the overlap between the states having momentums $\mathbf{k}$ and $\mathbf{k+q}$.
In the long wavelength limit $q\rightarrow 0$ (i.e. $q\ll \omega/v_{F}$) one may expand the denominator\cite{Phillips} and then
keep the first term of the Taylor series for both the difference in the Fermi functions and the energies ($E_{\mathbf{k}}-E_{\mathbf{k+q}}$).
Further, the overlap integral $F_{ll^{\prime}}$ is close to unity since the angle
between $\mathbf{k}$ and $\mathbf{k+q}$ is almost zero. Finally for the zero temperature case, the difference of the Fermi functions
will yield the factor $\delta (k-k_{F})$.
This leads to an approximate relation for the polarizability
\begin{equation}\label{Piapprox}
\Pi(\mathbf{q},\omega)\approx\frac{g_{d}k_{F}}{4\pi}\frac{q^{2}}{(\hbar\omega)^{2}}\frac{\partial E_{k}}{\partial k}|_{k=k_{F}}\,.
\end{equation}
 Taking into account the energy-momentum relation, Eq.~(\ref{eigenener}) and upon inserting Eq.~(\ref{Piapprox}) into Eq.~(\ref{dielPi}) one arrives at the plasmon dispersion relation
\begin{equation}\label{plasmdisp}
\omega_{\mathbf{q}}=\left(\frac{3g_{d}\hbar (v_{F}k_{F})^{3} e^2}{2\gamma_{1}^{2}\kappa}\right)^{\frac{1}{2}}\sqrt{q}\,.
\end{equation}
\noindent The Fermi wavevector can be calculated from the known relation $k_{F}=\sqrt{\pi n_{e}}$ which holds for all $2$D systems with isotropic energy dispersion.

The excitations of the electron gas can be represented by a scalar field previously described by Overhauser\cite{Overhauser} for the $3$D electron gas,
but here modified for the $2$D electron gas. The correction to the electron spectrum are calculated in a similar way as for the polaron problem
with the difference that a test charge interacts with plasmons. The interaction of an electron with plasmons was treated in our earlier work\cite{Krstaj_bil},
and the interaction term of the Hamiltonian is given by
\begin{equation}
H_{int}=\sum_{\mathbf{q}}\frac{V_{\mathbf{q}}}{\sqrt{\Omega}}\exp(i\mathbf{q}\cdot\mathbf{r})(a_{\mathbf{q}}+a_{-\mathbf{q}}^{\dagger})\,,
\end{equation}
\noindent where $a_{\mathbf{q}}$ and $a_{\mathbf{q}}^{\dagger}$ are electron annihilation and creation operators, respectively. Here the electron-plasmon interaction matrix element is\cite{PeetPol}
\begin{equation}\label{Vq1}
V_{\mathbf{q}}=\frac{2\pi e^{2}}{\sqrt{\Omega}\kappa q}\lambda_{\mathbf{q}}\,.
\end{equation}
 The value of $V_{\mathbf{q}}$ is determined using the $f$-sum rule applied to the case of interest. The derivation of the $f$-sum rule goes as follows.
 First, we note that the expectation value of the double commutator $\langle 0|[n_{-\mathbf{q}},[n_{\mathbf{q}},H]]|0\rangle$
 can be evaluated in two different ways\cite{Pines}. Here $n_{\mathbf{q}}$ is the electron density operator,
 \begin{equation}\label{nq}
 n_{\mathbf{q}}=\sum \lambda_{\mathbf{q}}\left(a_{\mathbf{q}}e^{i\mathbf{q}\cdot\mathbf{r}}+a_{\mathbf{q}}^{\dagger}e^{-i\mathbf{q}\cdot\mathbf{r}}\right)
 \end{equation}

  Then, it is known that the relation $\langle n|C|m\rangle=(E_{n}-E_{m})\langle n|A|m\rangle$
 holds for any commutator with the Hamiltonian, $C=[H,A]$. Second,
it can easily be proven that
\begin{equation}\label{Vq2}
\langle 0|[n_{-\mathbf{q}},[n_{\mathbf{q}},H]]|0\rangle=2\sum_{n}\hbar\omega_{n0}|\langle n|n_{\mathbf{q}}|0\rangle|^{2}\,,
\end{equation}
\noindent where $\hbar\omega_{n0}=E_{n}-E_{0}$. Then, the explicit evaluation of the double commutator yields
\begin{equation}\label{Vq3}
\sum_{n}\hbar\omega_{n0}|\langle n|n_{\mathbf{q}}|0\rangle|^{2}=N\frac{(\hbar v_{F})^{3}q^{3}}{\gamma_{1}^{2}}\,.
\end{equation}
Within the plasmon-pole approximation there is only one collective excitation for
each wave vector $\mathbf{q}$, so that one can put $\omega_{n0}=\omega_{\mathbf{q}}^{\prime}$ and taking into account Eq.~(\ref{nq}),
the sum rule reduces to
\begin{equation}\label{Vq4}
\hbar\omega_{\mathbf{q}}^{\prime}\lambda_{\mathbf{q}}^{2}=N\frac{(\hbar v_{F})^{3}q^{3}}{\gamma_{1}^{2}}\,.
\end{equation}
Here the quantity $\lambda_{q}^{\prime}=\sqrt{(\hbar v_{F})^{3}q^{3}/(\gamma_{1}^{2}\hbar\omega_{\mathbf{q}}^{\prime})}$ serves
as a small dimensionless parameter in the electron-plasmon interaction, and takes the value of about or less than $0.5$.
On inserting Eq.~(\ref{Vq4}) in Eq.~(\ref{Vq1}) one arrives at the following expression for the interaction matrix element
\begin{equation}\label{eqVq}
V_{\mathbf{q}}=\frac{2\pi e^{2}}{\kappa\gamma_{1}}\sqrt{\frac{(\hbar v_{F})^{3} qn_{e}}{\hbar\omega_{\mathbf{q}}^{\prime}}}\,,
\end{equation}
\noindent where $n_{e}=N/\Omega$ is the electron concentration. Note that $\omega_{\mathbf{q}}^{\prime}$
is not the bare plasmon frequency but is modified by the polarization of the electron gas. In order to investigate electron-plasmon interaction,
one should consider a test charge interacting with the plasmon modes. But this test charge introduces
a change in energy as a result of its interaction with the dielectric. In order to determine the value of the plasmon frequency
one needs the electron dielectric function. It can be shown that the static dielectric function within the
Random Phase Approximation can be approximated by the following relation\cite{dielSarma}
\begin{equation}\label{epsqE2}
\epsilon(q)=1+\frac{q_s}{q}\,,
\end{equation}
\noindent where $q_{s}$ is the screening wavevector\cite{SarmaRev} and given by $q_{s}=2\pi e^{2}/\kappa D_{0}$
where $D_{0}$ is the density of states of the \textit{ABC} stacked trilayer graphene,
\begin{equation}
D_{0}=\frac{2}{3}\left(\frac{g_{d}}{4\pi}\right)^{3/2}\cdot\frac{\gamma_{1}^{2}}{(\hbar v_{F})^{3}}\frac{1}{\sqrt{n_{e}}}\,.
\end{equation}

Here $g_{d}$ is the degeneracy factor that takes into account both spin and valley degrees of freedom. Then, the actual plasmon frequency is given by\cite{remotepol}
\begin{equation}\label{epsqE1}
\omega_{\mathbf{q}}^{\prime2}=\omega_{\mathbf{q}}^2\frac{\epsilon(q)}{\epsilon(q)-1}\,.
\end{equation}

Now, using second order perturbation theory, the correction to the energy spectrum is given by

\begin{equation}\label{corrE}
\Delta E_{0}(\mathbf{k})=-P\frac{1}{\Omega}\sum_{\mathbf{q}}\frac{|V_{\mathbf{q}}|^2}{\hbar\omega_{\mathbf{q}}+E_{0}(\mathbf{k}-\mathbf{q})-E_{0}(\mathbf{k})}\,,
\end{equation}
where $P(\cdot)$ stands for the principal value. Here, the cut-off value for the momentum $\mathbf{q}$ was taken to be $q_{c}=1/a_{0}$ where $a_{0}$ is the lattice constant.
This formula correspond to non-degenerate Rayleigh-Schr\"{o}dinger perturbation theory (RSPT). For certain values of the plasmon wavevector $\mathbf{q}$
a degeneracy occurs when $E_{0}(\mathbf{k})=\hbar\omega_{\mathbf{q}}+E_{0}(\mathbf{k}-\mathbf{q})$, and one should employ improved Wigner
Brillouin perturbation theory\cite{PeetPol} (IWBPT). The main idea behind this method is to ensure enhanced convergence when the denominator in Eq.~(\ref{corrE})
approaches zero, which is realized by adding the term $\Delta(\mathbf{k}) = \Delta E(\mathbf{k})-\Delta E_{0}(\mathbf{k})$ ($\Delta E_{0}(\mathbf{k})$ is evaluated within RSPT),

\begin{equation}\label{corrEIW}
\Delta E(\mathbf{k})=-P\sum_{\mathbf{q}}\frac{|V_{\mathbf{q}}|^2}{\hbar\omega_{\mathbf{q}}+E_{0}(\mathbf{k-q})-
E_{0}(\mathbf{k})-\Delta(\mathbf{k})}.
\end{equation}

This equation has to be solved self-consistently as $\Delta E$ appears on both sides of the equation.
Because of the isotropic nature of the spectrum, we have $E(\mathbf{k})=E(k)$. In the next section the value of $\Delta E(k)$ (within IWBPT)
will be calculated numerically for concrete values of the doping level, permittivity and other material parameters. As has been pointed out
elsewhere\cite{plasmaron3} the plasmon excitation in graphene of the Dirac sea remains pretty much
well defined even when it penetrates the interband particle-hole continuum. This is the consequence of the fact that the transitions near the bottom of
the interband particle-hole continuum have almost parallel wavevectors $\mathbf{k}$ and $\mathbf{k+q}$.
Thus, those transitions carry negligible charge-fluctuation weight.
A similar conclusion holds for trilayer graphene. In practice, the damping can be important for very large momentum $q$,
but then the contribution to the energy shift, i.e. to the integral in Eq.~(\ref{corrEIW}), is small.

\subsection{\textit{ABA} stacked trilayer graphene}
In this case multilayer graphene is stacked in the Bernal type where the sites in the first and the third layer coincide. This kind
of stacking is more common and can be realized by exfoliating natural graphite because it has virtually the same crystalline structure\cite{Bernal}.
The effective Hamiltonian obtained by a tight-binding model and taking into account only nearest-neighbor interaction is\cite{Ben1,BartP}

\begin{equation}
H_{ABA}=\hbar v_{F}\left[\begin{array}{ccc}
\boldsymbol{\sigma}\cdot\mathbf{k}+\delta^{\prime}I_{2} & \tau & 0  \\
\tau^{\dagger} & \boldsymbol{\sigma}\cdot\mathbf{k} & \tau^{\dagger} \\
0 & \tau & \boldsymbol{\sigma}\cdot\mathbf{k}-\delta^{\prime}I_{2}
\end{array}\right]\,,
\end{equation}
\noindent where $I_{2}$ is the $2\times2$ unit matrix and $\delta^{\prime}=\delta/(\hbar v_{F})$ is the externally
induced interlayer potential difference. The Hamiltonian is written in the basis of orbital eigenfunctions
\begin{equation}
\Psi=[\psi_{\alpha_{1}},\psi_{\beta_{1}},\psi_{\alpha_{2}},\psi_{\beta_{2}},\psi_{\alpha_{3}},\psi_{\beta_{3}}]^{T}\,,
\end{equation}
\noindent where the indices correspond to the different sublattices (\textit{A} or \textit{B}) of the three layers.
When the external potential is zero, $\delta=0$, the two blocks in the Hamiltonian correspond to a superimposed linear spectrum
(monolayer like) and a hyperbolic one (bilayer like) near the Dirac point. Then, electrons in \textit{ABA} stacked TLG may
propagate through two different modes, one monolayer-like and the other bilayer-like mode. The scattering between the two
modes is not allowed as long as the mirror symmetry of the three layers remains conserved. As for the plasmons in trilayer
graphene, one can envisage that the system in question is composed of monolayer and bilayer graphene and the dielectric function
can then be written in the $2\times2$ matrix form
\begin{equation}\label{dielMatr}
\epsilon(q,\omega)={\rm det}\left|I_{2}-\hat{v}(q)\cdot \hat{\Pi}(q)\right|\,,
\end{equation}
\noindent where
\begin{equation}
v_{ij}(q)=\frac{2\pi e^{2}}{q}e^{-|i-j|qd}\,.
\end{equation}
Here $d$ is the interlayer distance $d=3.42{\rm \AA}$, while $\Pi_{11}$ and $\Pi_{22}$ are the polarizability of single and bilayer graphene, respectively.
Here we assume that to leading order there is no direct coupling between the two modes so that $\Pi_{12}=\Pi_{21}=0$. Upon inserting $\Pi_{11}(\mathbf{q},\omega)=C_{1}q^{2}/\omega^{2}$
for single layer and $\Pi_{22}(\mathbf{q},\omega)=C_{2}q^{2}/\omega^{2}$ for bilayer in Eq.~(\ref{dielMatr}) one can
find the plasmon modes by determining the zeros of the dielectric function $\epsilon(\mathbf{q},\omega)$.
Here $C_{1}=g_{d}E_{F}/(4\pi \hbar^{2})$ and $C_{2}=2g_{d}E_{F}/(4\pi \hbar^{2})$. This leads to a biquadratic equation with respect to $\omega$,
\begin{equation}
\omega^{4}-(C_{1}+C_{2})\frac{2\pi e^{2}q}{\kappa}\omega^{2}+C_{1}C_{2}\frac{2\pi e^{2}q^{2}}{\kappa}(1-e^{-2qd})=0\,.
\end{equation}
However in practice $qd\ll1$ and $\exp(-2qd)\approx1-2qd$,
which yields the following expressions for the two plasmon modes, one optical like
\begin{subequations}
\begin{equation}
\omega_{op}=\sqrt{\frac{3e^{2}g_{d}E_{F}}{2\hbar^{2}\kappa}q}\,,
\end{equation}
and one acoustical like
\begin{equation}
\omega_{ac}=\sqrt{\frac{8\pi e^{2}}{\kappa}d}\cdot\frac{g_{d}E_{F}}{\hbar} q\,.
\end{equation}
\end{subequations}
The first mode has a square root dependence on the wavevector $q$ and the second one is linear in $q$.
\section{Numerical results}\label{results}
\begin{figure}[t]
\includegraphics[width=8cm]{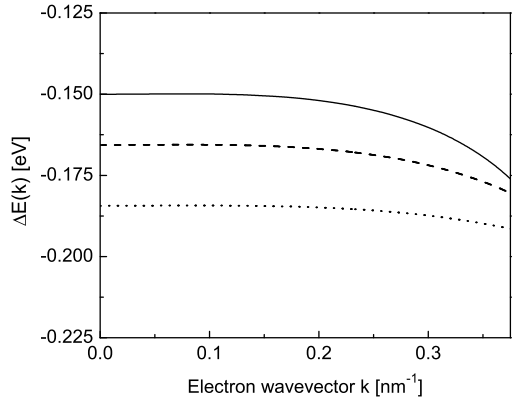}
\caption{\label{fig1N}The correction to the energy, $\Delta E(k)$, vs electron momentum $k$, in \textit{ABC} stacked trilayer graphene
for three values of the doping level $n_{e}=3\times10^{12}\,{\rm cm}^{-2}$ (solid curve), $5\times10^{12}\,{\rm cm}^{-2}$
(dashed curve) and $10^{13}\,{\rm cm}^{-2}$ (dotted curve).}
\end{figure}
We will present numerical calculations for doped trilayer graphene, with varying electron concentration.
\subsection{\textit{ABC} stacked trilayer graphene}
First, we give results for \textit{ABC} stacked trilayer graphene,
having cubic energy dispersion in the low-energy limit. Fig.~\ref{fig1N} shows the results for the energy correction
$\Delta E(\mathbf{k})$ for three values of the electron concentration: $n_{e}=3\times10^{12} \,{\rm cm}^{-2}$ (solid curve),
$5\times10^{12}\,{\rm cm}^{-2}$ (dashed curve) and $10^{13} \,{\rm cm}^{-2}$ (dotted curve). The value of background dielectric
constant was $\kappa_{s}=3.8$ that corresponds to SiO$_2$\cite{kappaSiO2}, and this value is approximately
the same for hexagonal boron nitride (h-BN) substrate\cite{kappahBN}.
As can be seen, the shift increases with the electron momentum,
and this dependence is more pronounced for lower electron concentrations.
The increase with $\mathbf{k}$ is more rapid than in the case of single layer graphene\cite{Krstajplasm}. Note that the explicit
dependence on the concentration is $V_{\mathbf{q}}\propto\sqrt{n_{e}}$ similar monolayer graphene,
but the interaction matrix element is also related to the doping level through the plasmon frequency.
The latter in single layer graphene is mainly proportional to $n_{e}^{1/4}$ while in trilayer graphene it has a more complicated
dependence which depends also on the stacking order.
Further, the effective plasmon frequency is modulated through the polarization of the surrounding electron gas, which depends on the
density of states.
On the other hand, the coupling parameter is a function of the carrier density $r_{s}=f(n_{e})$
(while in single layer graphene it is independent of $n_{e}$).

\begin{figure}[t]
\includegraphics[width=8cm]{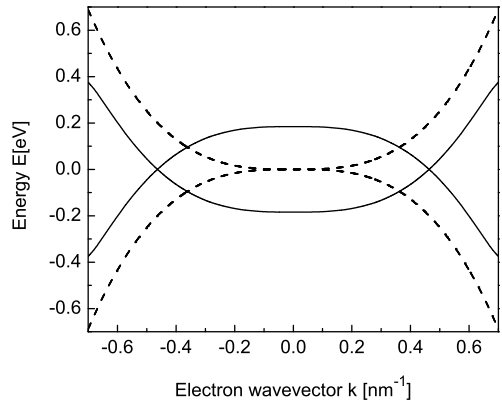}
\caption{\label{fig2N}Energy band structure of \textit{ABC} stacked trilayer graphene for electron concentration $n_{e}=10^{13} {\rm cm}^{-2}$, and
for $\kappa_{s}=3.8$. The dashed curves is for the case of zero electron-plasmon interaction, i.e. $n_{e}=0$.}
\end{figure}

 In contrast to the case of polarons in conventional semiconductors, here it is not straightforward to derive
any approximate analytical relation for $\Delta E(k)$ at small $k$. This is due to the fact that plasmons here
have a more complicated dispersion relation, and the fact that the interaction strength $V_{\mathbf{q}}$
depends on $\mathbf{q}$ in a non-trivial manner.
Thus we will treat Eq.~(\ref{corrE}) numerically and one may write for small $k$
\begin{equation}\label{apprdE}
\Delta E(k)=\Delta E(0)+\alpha k^3+\beta k^6\,.
\end{equation}
We fitted Eq.~(\ref{apprdE}) to our numerical results within the range $0<k<0.4 {\rm nm}^{-1}$. For instance, for $n_{e}=3\times10^{12}\,{\rm cm}^{-2}$
the fitting parameters are $\alpha=-1.98\times10^{-22}\,{\rm eVcm}$
and $\beta=-5.86\times10^{-42}\,{\rm eVcm}^{2}$.
\begin{figure}[t]
\includegraphics[width=8cm]{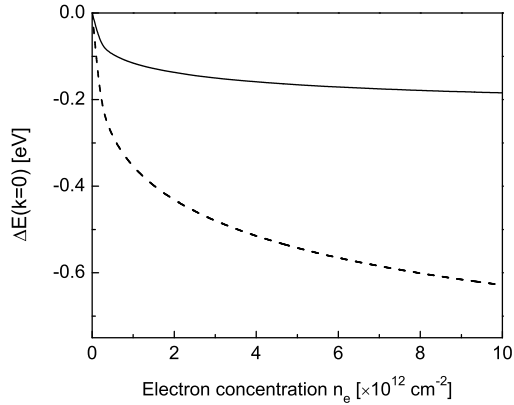}
\caption{\label{fig3N}The correction to the energy $\Delta E(0)$ for $k=0$ in \textit{ABC} stacked trilayer graphene, vs the electron concentration $n_{e}$,
for $\kappa_{s}=3.8$ (solid curve) and $\kappa_{s}=1$ (dashed curve).}
\end{figure}

Figure~\ref{fig2N} shows the energy band structure of \textit{ABC} stacked trilayer graphene at electron concentration $n_{e}=10^{13}{\rm cm}^{-2}$
($E_{F}=0.35{\rm eV}$) within the cubic
approximation. The dashed curve corresponds to unperturbed values in the absence of electron-plasmon interaction.
\begin{figure}[h]
\centering
\includegraphics[width=8cm]{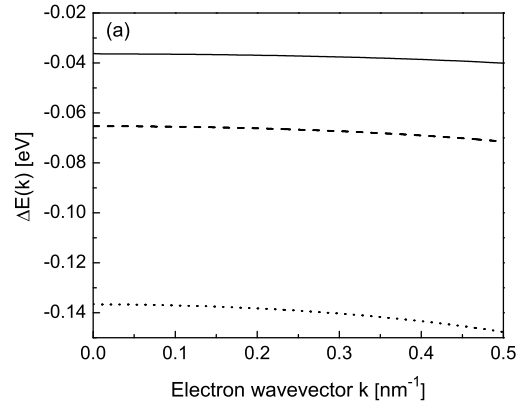}
\includegraphics[width=8cm]{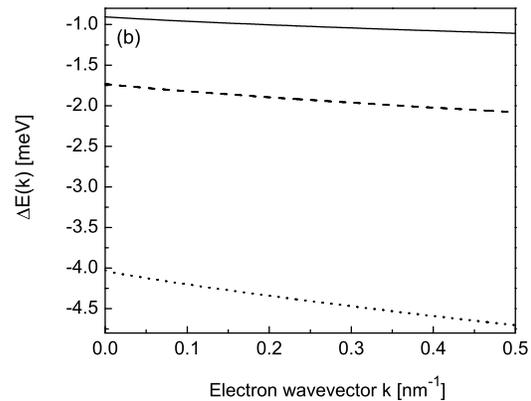}
\caption{The correction to the energy, $\Delta E(k)$, vs electron momentum $k$, in \textit{ABA} stacked trilayer graphene
for three values of the electron concentration $n_{e}=3\times10^{12}\,{\rm cm}^{-2}$ (solid curve), $5\times10^{12}\,{\rm cm}^{-2}$
(dashed curve) and $10^{13}\,{\rm cm}^{-2}$ (dotted curve). Dashed curves correspond to free-standing graphene.
The top (bottom) panel corresponds to the hyperbolic (linear) part of the energy spectrum.} \label{fig4N}
\end{figure}

In Fig.~\ref{fig3N} we present the result for the energy correction $\Delta E(0)$ at $k=0$, vs the electron
concentration $n_{e}$ in \textit{ABC} trilayer graphene. The solid curve corresponds to trilayer graphene on SiO$_2$ ($\kappa_{s}=3.8$), while
the dashed curve corresponds to free-standing graphene ($\kappa_{s}=1$). It can be seen that the absolute value of $\Delta E(0)$ increases with the electron concentration.
This is mainly due to the dependence of the matrix element $V_{\mathbf{q}}$ on the electron concentration $n_{e}$ (see Eqs.~(\ref{eqVq}) and (\ref{epsqE1})).
This relation is complicated since the plasmon frequency is modified through the polarization of the electron gas. On the other hand, the values for the case
of free-standing graphene are considerably higher because the effective dielectric constant is smaller and thus the interaction matrix element $V_{\mathbf{q}}$ is
larger. We found that the obtained results for the energy shift on the concentration can be fitted (solid curve)
(for $0<n_{e}<10^{13} {\rm cm}^{-2}$) to $\Delta E(0)=an_{e}^{\alpha}/(1+bn_{e}^{\gamma})$,
where $\alpha=0.55$, $\gamma=0.52$ and $a=-6.14\times10^{-8}$, $b=6.44\times10^{-5}$, and for the dashed curve
$\alpha=0.6$, $\gamma=0.48$ and $a=-5.78\times10^{-8}$, $b=2.76\times10^{-6}$
($n_{e}$ is expressed in ${\rm cm}^{-2}$ and $\Delta E(0)$ in ${\rm eV}$).
\subsection{\textit{ABA} stacked trilayer graphene}
Next, we consider the case of \textit{ABA} stacked trilayer graphene. As has already been mentioned this kind of stacking is an arrangement where
the sites in the first and the third layer coincide. Since the spectrum consists of a hyperbolic and a linear part, we will consider them
separately. Fig.~\ref{fig4N} shows the correction
to the energy $\Delta E(k)$ for three values of the electron concentration $n_{e}=3\times10^{12}\,{\rm cm}^{-2}$ (solid curve), $5\times10^{12}\,{\rm cm}^{-2}$
(dashed curve) and $10^{13}\,{\rm cm}^{-2}$ (dotted curve). The background dielectric constant was taken that of SiO$_2$, $\kappa_{s}=3.8$.
The top and the bottom panel correspond to the hyperbolic and linear part of the energy spectrum, respectively.
The shift is larger for higher electron concentration as expected, and lies in the range $30\div150 {\rm meV}$ for the hyperbolic part and $1\div5 {\rm meV}$ for the linear
part of the energy spectrum.
The dispersion is less pronounced than in the case of \textit{ABC} trilayer graphene
which is the consequence of the different energy band structure and plasmon dispersion. Note that in the case of the energy correction to the linear part,
the shape of curves have different convexity than in the first case. The values of $\Delta E(k)$ in Fig.~\ref{fig4N}(a) are lower than in the case of bilayer graphene\cite{Krstaj_bil}
for all three electron concentration. The same holds true for the linear part of the spectrum, Fig.~\ref{fig4N}(b), where the values are considerably lower\cite{Krstajplasm}.
One of the reasons is that the Fermi energy is determined by the electron concentration in trilayer structure as a whole, which is distributed over two bands
 and has in general lower values.
\begin{figure}[h]
\includegraphics[width=9cm]{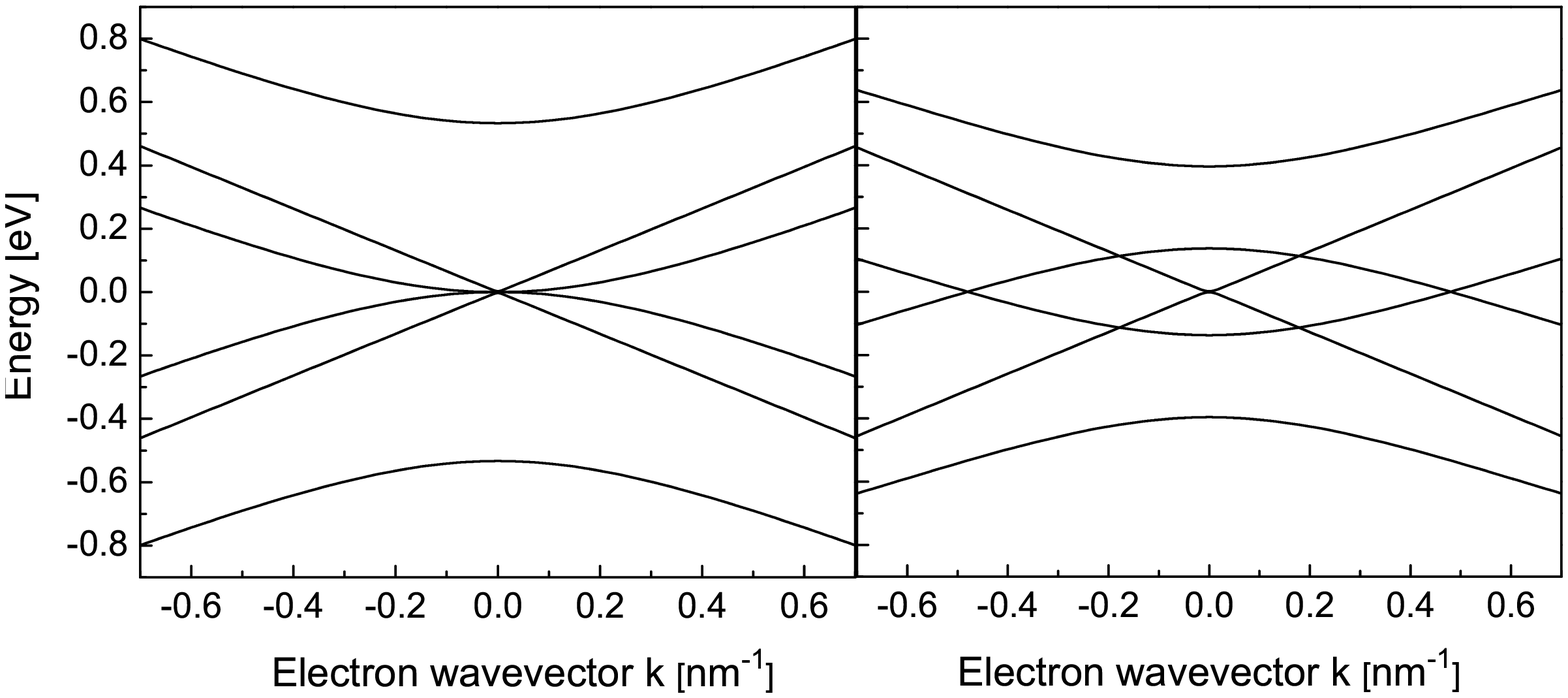}
\caption{Energy band structure of \textit{ABA} stacked trilayer graphene for electron concentration $n_{e}=10^{13}{\rm cm}^{-2}$, and $\kappa_{s}=3.8$.
Left panel corresponds to the absence of electron-plasmon interaction, while the right panel is
for the case when electron-plasmon interaction is taken into account.} \label{fig5N}
\end{figure}

Figure~\ref{fig5N} shows the energy band structure of \textit{ABA} stacked trilayer graphene when the interaction between electrons and plasmons
are taken into account (right panel), and in the absence of this interaction (left panel). The electron concentration is taken to be $n_{e}=10^{13}{\rm cm}^{-2}$
($E_{F}=0.6{\rm eV}$).
It consists of two groups of branches one belonging
to the linear part of the spectrum and the second to the hyperbolic part. The linear part is barely shifted from the unperturbed part, since the
values of the energy shift are of order of several meV (see Fig.~\ref{fig4N}(b)).
\begin{figure}[t]
\centering
\includegraphics[width=8cm]{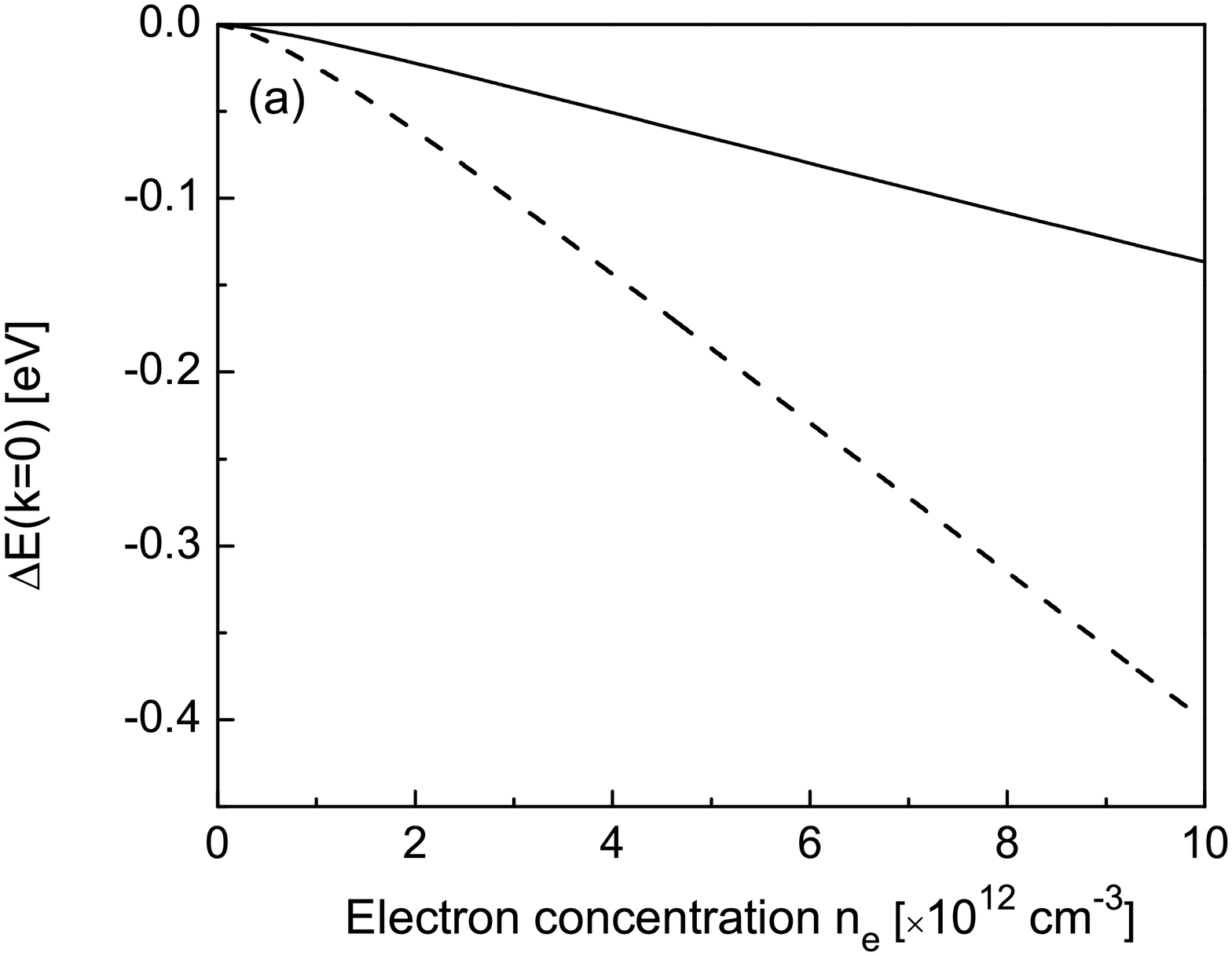}
\includegraphics[width=8cm]{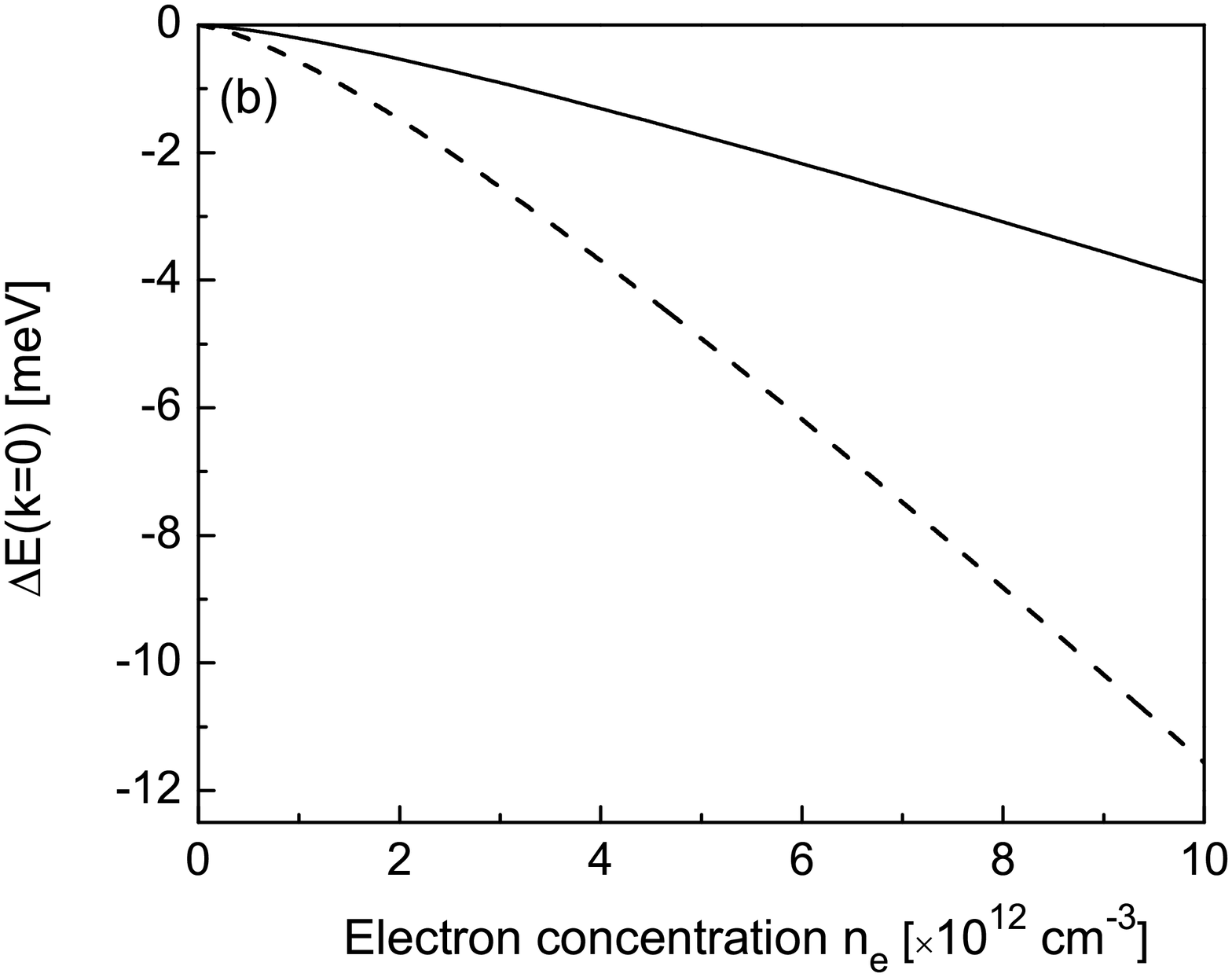}
\caption{The correction to the energy, $\Delta E(0)$, for zero momentum $k=0$ vs electron concentration $n_{e}$,
in \textit{ABA} stacked trilayer graphene. Dashed curves correspond to free-standing graphene and the solid curve to graphene on SiO$_2$.
The top (bottom) panel corresponds to the hyperbolic (linear) part of the energy spectrum.} \label{fig6N}
\end{figure}

Then in Fig.~\ref{fig6N} we give the correction to the energy at zero momentum, but as a function of the electron concentration.
The top and the bottom panel correspond to the hyperbolic and linear part of the energy spectrum, respectively. The solid curves correspond
to trilayer graphene on SiO$_2$ substrate while the dashed curves correspond to free-standing graphene.
The energy correction $\Delta E(0)$ increases with the electron concentration and exhibits almost a linear dependence for the hyperbolic part.
The values for free-standing graphene are larger in absolute sense since then the interaction matrix element $V_{\mathbf{q}}$ is larger.
The value of $\Delta E(0)$ can be fitted (solid curve) to $\Delta E(0)=c_{1}n_{e}+c_{2}n_{e}^{2}$ where $c_{1}=-1.19\times10^{-14} {\rm eVcm^{2}}$
and $c_{2}=-1.99\times10^{-28}{\rm eVcm^{4}}$ for the hyperbolic part, while for the
linear part $\Delta E(0)=c_{1}n_{e}+c_{2}n_{e}^{2}$ where $c_{1}=-2.72\times10^{-16}{\rm eVcm^{2}}$ and $c_{2}=-1.38\times10^{-29}{\rm eVcm^{4}}$.
The dashed curves can be fitted with the following coefficients $c_{1}=-3.27\times10^{-14}{\rm eVcm^{2}}$ and $c_{2}=-8.0\times10^{-28}{\rm eVcm^{4}}$
(linear part) and $c_{1}=-7.58\times10^{-16}{\rm eVcm^{2}}$ and $c_{2}=-4.19\times10^{-29}{\rm eVcm^{4}}$ (hyperbolic part).
Note that values of the energy shift are smaller than in cases of single monolayer\cite{Krstajplasm} and bilayer graphene\cite{Krstaj_bil} since
the Fermi energy has lower values as the electron concentration is distributed over the two bands.

\section{Conclusion}\label{summary}
In this work we investigated the interaction between an electron and plasmons, i.e. the collective excitation of the electron gas,
in trilayer graphene by employing a field-theoretical approach. We considered both \textit{ABC} and \textit{ABA} stacking order
which differ in both their energy spectrum and their plasmon dispersion. The motivation behind the present study are the increased interest
in transport and optical properties of trilayer
graphene\cite{ABC1,ABC2}. The interaction between electrons and plasmons is modeled by applying
the Overhauser approach \cite{Overhauser} to the case of interest. We evaluated the energy correction, that is
the shift in the energy spectrum as a result of this interaction. Second order perturbation theory was employed in
order to determine the energy of the plasmaron, which is a bound state of an electron with a cloud of plasmons, and serves as a composite particle.

First we evaluated the correction to the energy as a result of the interaction
between electron and plasmons, for the cases with \textit{ABC} and \textit{ABA} stacking order.
The shift is appreciable and lies in the range
of $150\div200 {\rm meV}$ for \textit{ABC} stacked trilayer graphene. As for \textit{ABA} stacked trilayer graphene the
energy correction should be evaluated for the hyperbolic and linear part of the spectrum and amounts to $30\div150 {\rm meV}$ and
$1\div5 {\rm meV}$, respectively for graphene on SiO$_2$ with its dielectric constant being $\kappa_{s}=3.8$.
The shift, of course depends and rises in absolute value with the electron concentration and electron wavevector.

Further, we investigated the influence of the doping level on the shift $\Delta E(0)$,
and it is shown that it increases with $n_{e}$ which is more pronounced than in the case of single layer graphene\cite{Krstajplasm}.
The difference with single layer graphene lies in the actual dependence of the interaction strength $V_{\mathbf{q}}$
on the electron concentration. The energy correction for \textit{ABC} and \textit{ABA} stacking order (only the hyperbolic part)
has the same order of magnitude as recently calculated for bilayer graphene\cite{Krstaj_bil}.

At the end we discuss available experimental data related to the electronic structure of trilayer graphene. To our knowledge,
there exists currently only one experimental investigation\cite{exp1} of the electronic structure of trilayer and bilayer graphene on Ru(0001) using
selected-area angle-resolved photoelectron spectroscopy (micro-ARPES). However, it was determined in that work that
there is a strong coupling between the first graphene layer and the adjacent metal (Ru) that disrupts the graphene bands near
the Fermi energy. This perturbation vanishes rapidly with the addition of subsequent graphene sheets. Therefore, trilayer graphene on Ru
behaves like free-standing bilayer graphene. Consequently, the experimental data of Ref.~\onlinecite{exp1} are not related to
our results. We hope that new experimental data will emerge in literature in near future, so that one may test and verify
the theoretical results given in this work.
\acknowledgments{}
This work was supported by the Flemish Science Foundation (FWO-Vl), the ESF-EuroGRAPHENE project CONGRAN
and by the Serbian Ministry of Education and Science, within
the project No. TR~32008.

\end{document}